\documentstyle[12pt,epsf,psfig,graphicx]{article}
\begin{document}

\title{A recent appreciation of the singular dynamics at the edge of chaos}
\author{E. Mayoral and 
A. Robledo\\Instituto de F\'{i}sica,
Universidad Nacional Aut\'{o}noma de M\'{e}xico, \\Apartado Postal 20-364,
M\'{e}xico 01000 D.F., Mexico\\\texttt{robledo@fisica.unam.mx}}
\maketitle

\begin{abstract}
We study the dynamics of iterates at the transition to chaos in the logistic
map and find that it is constituted by an infinite family of Mori's $q$%
-phase transitions. Starting from Feigenbaum's $\sigma $ function for the
diameters ratio, we determine the atypical weak sensitivity to initial
conditions $\xi _{t}$ associated to each $q$-phase transition and find that
it obeys the form suggested by the Tsallis statistics. The specific values
of the variable $q$ at which the $q$-phase transitions take place are
identified with the specific values for the Tsallis entropic index $q$ in
the corresponding $\xi _{t}$. We describe too the bifurcation gap induced by
external noise and show that its properties exhibit the characteristic
elements of glassy dynamics close to vitrification in supercooled liquids,
e.g. two-step relaxation, aging and a relationship between relaxation time
and entropy.
\end{abstract}
\section{Introduction}

The logistic equation was introduced in 1845 by the Belgian mathematician
and sociologist Pierre-Fran\c{c}ois Verhulst to model the growth of
populations limited by finite resources \cite{verhulst1}. The discrete time
variable version of Verhulst's growth law, the logistic map, has become a
foundation stone for the theory of nonlinear dynamics. The logistic map is
the archetypal example of how the use as starting point of simple non-linear
discrete maps have often led to significant developments in the theory of
non-linear dynamical systems \cite{schuster1}. The discovery of the
universal properties associated to the renowned period-doubling and
intermittency routes to chaos displayed by the logistic map, similar to
those of conventional critical phenomena in statistical physics, triggered,
about three decades ago, an upsurge of activity in the field and now both
routes, as well as many other remarkable features displayed by the logistic
map, are well understood.

Here we might argue that at the present time the logistic map is again
becoming a prototypical model system. This time for the assessment of the
validity and understanding of the reasons for applicability of the
nonextensive generalization of the Boltzmann Gibbs (BG) statistical
mechanics \cite{tsallis0}, \cite{tsallis1}. This is a formalism assumed to
be appropriate for circumstances where the system is out of the range of
validity of the canonical BG theory. And these circumstances are believed in
some cases to be a breakdown in the chain of increasing randomness from
non-ergodicity to completely developed phase-space mixing.

The logistic map contains infinite families of critical attractors at which
the ergodic and mixing properties breakdown. These are the tangent
bifurcations and the accumulation point(s) of the pitchfork bifurcations,
the so-called onset of chaos \cite{schuster1}. At each of the map critical
attractors the Lyapunov coefficient $\lambda _{1}$ vanishes, and the
sensitivity to initial conditions $\xi _{t}$ for large iteration time $t$
ceases to obey exponential behavior, exhibiting instead power-law or faster
than exponential behavior. The pitchfork bifurcations are also critical
attractors at which the negative Lyapunov coefficient of periodic orbits
goes to cero. There are other attractors at which the Lyapunov coefficient
diverges to minus infinity, where there is faster than exponential
convergence of orbits. These are the superstable attractors located between
successive pitchfork bifurcations and their accumulation point is also the
onset of chaos.

Here we review briefly specific and rigorous results on the dynamics
associated to the critical attractors of the logistic map, or of its
generalization to non-linearity of order $z>1$, $f_{\mu }(x)=1-\mu \left|
x\right| ^{z}$,$\;-1\leq x\leq 1$, $0\leq \mu \leq 2$. (The phase space
variable is $x$, the control parameter is $\mu $ and the conventional
logistic map corresponds to $z=2$). Our results relate to the anomalous
sensitivity to initial conditions at the onset of chaos, the associated
spectrum of Tsallis $q$-Lyapunov coefficients, and the relationship of these
with the Mori $q$-phase transitions \cite{mori1}, one of which was
originally observed numerically for the Feigenbaum attractor \cite{mori1} 
\cite{mori2}. In particular, we identify the Mori singularities in the
Lyapunov spectra with the appearance of special values for the Tsallis
entropic index $q$. As the properties of the logistic map are very familiar
and well understood it is of interest to see how previous knowledge fits in
with the new perspective.

Tsallis suggested \cite{tsallis2} that for critical attractors $\xi _{t}$
(defined as $\xi _{t}(x_{0})\equiv \lim_{\Delta x_{0}\to 0}(\Delta
x_{t}/\Delta x_{0})$ where $\Delta x_{0}$ is the initial separation of two
orbits and $\Delta x_{t}$ that at time $t$), has the form 
\begin{equation}
\xi _{t}(x_{0})=\exp _{q}[\lambda _{q}(x_{0})\ t]\equiv [1-(q-1)\lambda
_{q}(x_{0})\ t]^{-1/(q-1)},  \label{sensitivity1}
\end{equation}
that yields the customary exponential $\xi _{t}$ with $\lambda _{1}$ when $%
q\rightarrow 1$. In Eq. (\ref{sensitivity1}) $q$ is the entropic index and $%
\lambda _{q}$ is the $q$-generalized Lyapunov coefficient; $\exp
_{q}(x)\equiv [1-(q-1)x]^{-1/(q-1)}$ is the $q$-exponential function.
Tsallis also suggested \cite{tsallis2} that the Pesin identity $K_{1}=$ $%
\lambda _{1}$ (where the rate of entropy production $K_{1}$ is given by $%
K_{1}t=S_{1}(t)-S_{1}(0)$, $t$ large, and $S_{1}=-\sum_{i}p_{i}\ln p_{i}$)
would be generalized to $K_{q}=$ $\lambda _{q}$, where the $q$-generalized
rate of entropy production $K_{q}$ is defined via $K_{q}t=S_{q}(t)-S_{q}(0)$%
, $t$ large, and where 
\begin{equation}
S_{q}\equiv \sum_{i}p_{i}\ln _{q}\left( \frac{1}{p_{i}}\right) =\frac{%
1-\sum_{i}^{W}p_{i}^{q}}{q-1}  \label{tsallisentropy1}
\end{equation}
is the Tsallis entropy; $\ln _{q}y\equiv (y^{1-q}-1)/(1-q)$ is the inverse
of $\exp _{q}(y)$.

To check on the above, we have analyzed recently \cite{robledo1}-\cite
{robledo4} both the pitchfork and tangent bifurcations and the onset of
chaos of the logistic map and found that indeed the Tsallis suggestions hold
for these critical attractors, though with some qualifications. For the case
of the tangent bifurcation it is important to neglect the feedback mechanism
into the neighborhood of the tangency to avoid the crossover out of the $q$%
-exponential form for $\xi _{t}$. As we explain below, for the onset of
chaos there is a multiplicity of $q$-indexes, appearing in pairs $q_{j}$,
and $Q_{j}=2-q_{j}$, $j=0,1,...$, and a spectrum of $q$-Lyapunov
coefficients $\lambda _{q_{j}}^{(k,l)}$ and $\lambda _{Q_{j}}^{(k,l)}$ for
each $q_{j}$ and $Q_{j}$, respectively. (The superindex $(k,l)$ refers to
starting and final trajectory positions). The dynamics of the attractor
confers the $q$-indexes a decreasing order of importance. Retaining only the
dominant indexes $q_{0}$ and $Q_{0}$ yields a quite reasonable description
of the dynamics and considering the next few leading indexes provides a
satisfactorily accurate account.

Although in brief, we also describe our finding \cite{robglass1} that the
dynamics at the noise-perturbed edge of chaos in logistic maps is analogous
to that observed in supercooled liquids close to vitrification. The three
major features of glassy dynamics in structural glass formers, two-step
relaxation, aging, and a relationship between relaxation time and
configurational entropy, are displayed by orbits with vanishing Lyapunov
coefficient. The known properties in control-parameter space of the
noise-induced bifurcation gap play a central role in determining the
characteristics of dynamical relaxation at the chaos threshold.

\section{Critical attractors in the logistic map}

For our purposes it is convenient to recall some essentials of logistic map
properties. The accumulation point of the period doublings and also of the
chaotic band splittings is the Feigenbaum attractor that marks the threshold
between periodic and chaotic orbits, at $\mu _{\infty }(z)$, with $\mu
_{\infty }=1.40115...$ when $z=2$ . The locations of period doublings, at $%
\mu =\mu _{n}<\mu _{\infty }$, and band splittings, at $\mu =\widehat{\mu }%
_{n}>\mu _{\infty }$, obey for large $n$ power laws of the form $\mu
_{n}-\mu _{\infty }\sim \delta (z)^{-n}$ and $\mu _{\infty }-\widehat{\mu }%
_{n}\sim \delta (z)^{-n}$, with $\delta =0.46692...$ when $z=2$, which is
one of the two Feigenbaum's universal constants. For our use below we recall
also the sequence of parameter values $\overline{\mu }_{n}$ employed to
define the diameters $d_{n}$ of the bifurcation forks that form the
period-doubling cascade sequence. At $\mu =$ $\overline{\mu }_{n}$ the map
displays a `superstable' periodic orbit of length $2^{n}$ that contains the
point $x=0$. For large $n$ the distances to $x=0$, of the iterate positions
in such $2^{n}$-cycle that are closest to $x=0$, $d_{n}\equiv f_{\overline{%
\mu }_{n}}^{(2^{n-1})}(0)$, have constant ratios $d_{n}/d_{n+1}=-\alpha (z)$%
; $\alpha =2.50290...$ when $z=2$, which is the second of the Feigenbaum's
constants. A set of diameters with scaling properties similar to those of $%
d_{n}$ can also be defined for the band splitting sequence \cite{schuster1}.
Other diameters in the $2^{n}$-supercycles are defined as the distance of
the $m$th element $x_{m}$ to its nearest neighbor $f_{\overline{\mu }%
_{n}}^{(2^{n-1})}(x_{m})$. That is 
\begin{equation}
d_{n,m}\equiv f_{\overline{\mu }_{n}}^{(m+2^{n-1})}(0)-f_{\overline{\mu }%
_{n}}^{(m)}(0),\ m=0,1,2,...,  \label{diameters1}
\end{equation}
with $d_{n,0}=d_{n}$. Feigenbaum \cite{feigenbaum1} constructed the
auxiliary function 
\begin{equation}
\sigma _{n}(m)=\frac{d_{n+1,m}}{d_{n,m}}  \label{sigmafunction1}
\end{equation}
to quantify the rate of change of the diameters and showed that it has
finite (jump) discontinuities at all rationals, as can be seen by
considering the variable $y=m/2^{n+1}$ with $n$ large (and therefore
omitting the subindex $n$). One obtains \cite{feigenbaum1} \cite{schuster1} $%
\sigma (0)=-1/\alpha $, but $\sigma (0^{+})=1/\alpha ^{z}$, and through the
antiperiodic property $\sigma (y+1/2)=-\sigma (y)$, also $\sigma
(1/2)=1/\alpha $, but $\sigma (1/2+0^{+})=-1/\alpha ^{z}$. Other
discontinuities in $\sigma (y)$ appear at $y=1/4,1/8,\ $etc. As these
decrease rapidly in most cases it is only necessary to consider the first
few.

An important factor of our work is that the sensitivity to initial
conditions $\xi _{t}(x_{0})$ can be evaluated for trajectories within the
Feigenbaum attractor via consideration of the discontinuities of $\sigma
_{n}(m)$. If the initial separation is chosen to be a diameter $\Delta
x_{0}=d_{n,m}$ and the final time $t$ is chosen to have the form $t=2^{n}-1$%
, then $\Delta x_{t}=d_{n,m+2^{n}-1}$, and $\xi _{t}(x_{0})$ can be written 
\cite{robestela1} as 
\begin{equation}
\xi _{t}(m)\simeq \left| \frac{\sigma _{n}(m-1)}{\sigma _{n}(m)}\right|
^{n},\ t=2^{n}-1,\ n\quad \textrm{large}.  \label{sensitivity2}
\end{equation}
For clarity of presentation, we shall only use absolute values of positions
so that the dynamics of iterates do not carry information on the
self-similar properties of ``left'' and ``right'' symbolic dynamic sequences 
\cite{schuster1}. This choice does not affect results on the sensitivity to
initial conditions.

\section{Mori's $q$-phase transitions in the logistic map}

During the late 1980's Mori and coworkers developed a comprehensive
thermodynamic formalism to characterize drastic changes at bifurcations and
at other singular phenomena in low dimensional maps \cite{mori1}. The
formalism was also adapted to the study of critical chaotic attractors and
was illustrated by considering the specific case of the onset of chaos in
the logistic map \cite{mori1} \cite{mori2} \cite{politi1}. For critical
attactors the scheme involves the evaluation of fluctuations of the
generalized finite-time Lyapunov coefficient 
\begin{equation}
\lambda (t,x_{0})=\frac{1}{\ln t}\sum_{i=0}^{t-1}\ln \left| \frac{df_{\mu
_{\infty }}(x_{i})}{dx_{i}}\right| ,\quad t\gg 1.  \label{lyafinitetime}
\end{equation}
Notice the replacement of the customary $t$ by $\ln t$ above, as the
ordinary Lyapunov coefficient 
\begin{equation}
\lambda _{1}(x_{0})=\lim_{t\rightarrow \infty }\frac{1}{t}%
\sum_{i=0}^{t-1}\ln \left| \frac{df_{\mu _{\infty }}(x_{i})}{dx_{i}}\right|
\label{lyaordinary}
\end{equation}
vanishes at $\mu _{\infty }$, $t\rightarrow \infty $.

The probability density distribution for the values of $\lambda $, $%
P(\lambda ,t)$, is written in the form 
\begin{equation}
P(\lambda ,t)=t^{-\psi (\lambda )}P(0,t),\quad t\gg 1,
\label{lyadistribution}
\end{equation}
where $\psi (\lambda )$ is a concave spectrum of the fluctuations of $%
\lambda $ with minimum $\psi (0)=0$ and is obtained as the Legendre
transform of the 'free energy' function $\phi (q)$, defined as 
\begin{equation}
\phi (q)\equiv -\lim_{t\rightarrow \infty }\frac{1}{\ln t}\ln Z(t,q),
\label{free energy phi}
\end{equation}
where $Z(t,q)$ is the dynamic partition function 
\begin{equation}
Z(t,q)\equiv \int d\lambda \ P(\lambda ,t)\ t^{-(q-1)\lambda }.
\label{partition function}
\end{equation}
The 'coarse-grained' function of generalized Lyapunov coefficients $\lambda
(q)$ is given by $\lambda (q)\equiv d\phi (q)/dq$ and the variance $v(q)$ of 
$P(\lambda ,t)$ by $v(q)\equiv d\lambda (q)/dq$ \cite{mori1} \cite{mori2}.
Notice the special weight $t^{-(q-1)\lambda }$ in the partition function $%
Z(t,q)$ as this shapes the quantities derived from it. These functions are
the dynamic counterparts of the Renyi dimensions $D_{q}$ and the spectrum $f(%
\widetilde{\alpha })$ that characterize the geometric structure of the
attractor.

A ''$q$-phase'' transition is indicated by a section of linear slope $%
m_{c}=1-q_{c}$ in the spectrum (free energy) $\psi (\lambda )$, a
discontinuity at $q_{c}$ in the Lyapunov function (order parameter) $\lambda
(q)$, and a divergence at $q_{c}$ in the variance (susceptibility) $v(q)$.
For the onset of chaos at $\mu _{\infty }(z=2)$ a single $q$-phase
transition was numerically determined \cite{mori1} \cite{mori2} \cite
{politi1} and found to exhibit a value close to $m_{c}=-(1-q_{c})\simeq -0.7$%
; arguments were provided for this value to be $m_{c}=-(1-q_{c})=-\ln 2/\ln
\alpha =-0.7555...$ Our analysis described below shows that the older
results give a broad picture of the dynamics at the Feigenbaum attractor and
that actually an infinite family of $q$-phase transitions of decreasing
weights take place at $\mu _{\infty }$.

\section{Tsallis dynamics at the edge of chaos}

By taking as initial condition $x_{0}=0$ we found that the resulting orbit
consists of trajectories made of intertwined power laws that asymptotically
reproduce the entire period-doubling cascade that occurs for $\mu <\mu
_{\infty }$ \cite{robledo2} \cite{robledo4}. This orbit captures the
properties of the so-called 'superstable' orbits at $\overline{\mu }_{n}<$ $%
\mu _{\infty }$, $n=1,2,...$ \cite{schuster1} (see Fig. 1), and can be used
as reference to read all other orbits within the attractor. At $\mu _{\infty
}$ the Lyapunov coefficient $\lambda _{1}$ vanishes and in its place there
appears a spectrum of $q$-Lyapunov coefficients $\lambda _{q}^{(k,l)}$. This
spectrum was originally studied in Refs. \cite{politi1} and \cite{mori2} and
our recent interest has been to study its properties in more detail to
examine their relationship with the Tsallis statistics. Recent analytical
results about the $q$-Lyapunov coefficients and the $q$-generalized Pesin
identity are given in Refs. \cite{robledo2} and \cite{robledo4}.

\begin{figure}[htb]
\setlength{\abovecaptionskip}{0pt} 
\centering
\includegraphics[width=9cm 
,angle=-90]{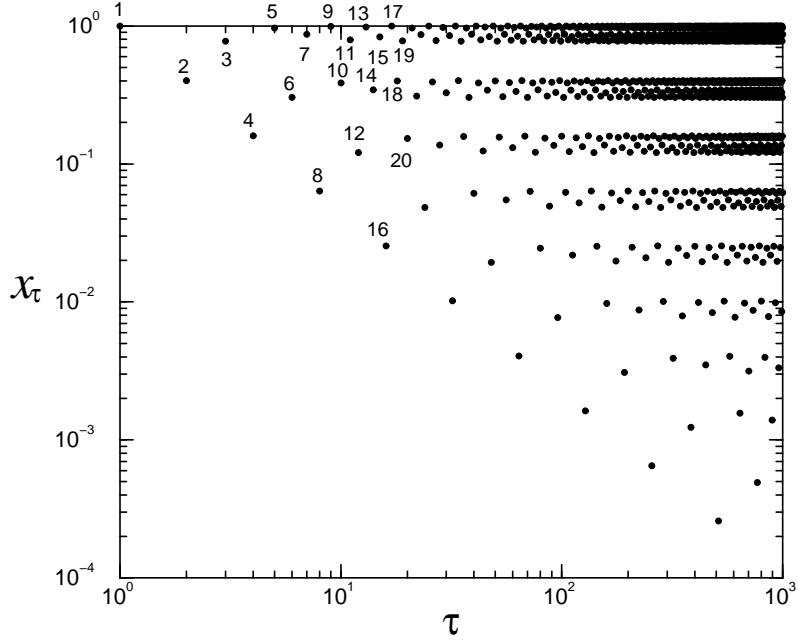}
\caption{ Absolute values of positions in logarithmic scales of iterations $%
\tau $ for a trajectory at $\mu _{\infty }$ with initial condition $x_{0}=0$%
. The numbers correspond to iteration times. The power-law decay of the time
subsequences can be clearly appreciated.}
\label{fig1}
\end{figure}

Now, consider a first approximation to the function $\sigma _{n}(m)$ for $n$
large, applicable to general non-linearity $z>1$. This is to assume that
half of the diameters scale as $\alpha $ (as in the most sparse region of
the attractor)\ while the other half scale as $\alpha _{0}=\alpha ^{z}$ (as
in the most crowded region of the attractor). This approximation captures
the effect on $\xi _{t}$ of the most dominant trajectories within the
attractor. With these two scaling factors $\sigma _{n}(m)$ becomes the
periodic step function 
\begin{equation}
\frac{1}{\sigma _{n}(m)}=\left\{ 
\begin{array}{l}
\quad \alpha _{0}=\alpha ^{z}, \\ 
\quad \alpha , \\ 
-\alpha _{0}=-\alpha ^{z}, \\ 
-\alpha \ ,
\end{array}
\right. 
\begin{array}{l}
\qquad 0<m\leq 2^{n-1}, \\ 
\quad 2^{n-1}<m\leq 2\cdot 2^{n-1}, \\ 
2\cdot 2^{n-1}<m\leq 3\cdot 2^{n-1}, \\ 
3\cdot 2^{n-1}<m\leq 4\cdot 2^{n-1},...
\end{array}
\label{sigmafunction2}
\end{equation}
Use of Eqn. (\ref{sigmafunction2}) into Eqn. (\ref{sensitivity2}) for the
sensitivity $\xi _{t}(m)$ yields the result 
\begin{equation}
\xi _{t}(m)=\left\{ 
\begin{array}{l}
\alpha ^{-(z-1)n}, \\ 
\alpha ^{(z-1)n}\ ,
\end{array}
\begin{array}{l}
m=(2k+1)2^{n-1}, \\ 
m=(2k+2)2^{n-1},
\end{array}
\right.   \label{sensitivity3}
\end{equation}
where $k=0,1,...$ and where the final observation time is of the form $%
t=l2^{n}-1$, $l=1,2,...$. With the introduction of the total time variable $%
\tau \equiv m+1+t$ Eq. (\ref{sensitivity3}) can be rewritten in terms of the 
$q$-exponential functions 
\begin{equation}
\xi _{\tau }(m)=\left\{ 
\begin{array}{l}
\lbrack 1+(1-q_{0})\lambda _{q_{0}}^{(k,l)}\tau ]^{1/(1-q_{0})}\ , \\ 
\lbrack 1+(1-Q_{0})\lambda _{Q_{0}}^{(k,l)}\tau ]^{1/(1-Q_{0})},
\end{array}
\begin{array}{l}
m=(2k+2)2^{n-1}, \\ 
m=(2k+1)2^{n-1},
\end{array}
\right.   \label{sensitivity4}
\end{equation}
where 
\begin{equation}
q_{0}=1-\frac{\ln 2}{\ln \alpha _{0}/\alpha }=1-\frac{\ln 2}{(z-1)\ln \alpha 
},  \label{q0index}
\end{equation}
\begin{equation}
\lambda _{q_{0}}^{(k,l)}=\frac{(z-1)\ln \alpha }{(k+2l+1)\ln 2},
\label{Lyaq0}
\end{equation}
\begin{equation}
Q_{0}=1+\frac{\ln 2}{\ln \alpha _{0}/\alpha }=1+\frac{\ln 2}{(z-1)\ln \alpha 
},  \label{Q0index}
\end{equation}
\begin{equation}
\lambda _{Q_{0}}^{(k,l)}=-\frac{2(z-1)\ln \alpha }{(2k+4l+1)\ln 2}.
\label{LyaQ0}
\end{equation}
Notice that $Q_{0}=2-q_{0}$. For $z=2$ one obtains $Q_{0}\simeq 1.7555$ and $%
q_{0}\simeq 0.2445$, this latter value agrees with that obtained in several
earlier studies \cite{tsallis2} \cite{tsallis3} \cite{robledo2} \cite
{robledo4}. The two scales considered describe correctly only trajectories
that start at the most sparse region of the multifractal ($x\simeq 0$) and
terminate at its most crowded region ($x\simeq 1$), or the inverse. (This is
why we obtain the two conjugate values $q$ and $Q=2-q$, as the inverse of
the $q$-exponential function satisfies $\exp _{q}(y)=1/\exp _{2-q}(-y)$).
The vertical lines in Fig. 2a represent the ranges of values obtained when $%
z=2$ for $\lambda _{q_{0}}^{(k,l)}$ and $\lambda _{Q_{0}}^{(k,l)}$. See Ref. 
\cite{robestela1} for more details.

\begin{figure}[htb]
\setlength{\abovecaptionskip}{0pt}
 \centering
\includegraphics[width=9cm 
,angle=0]{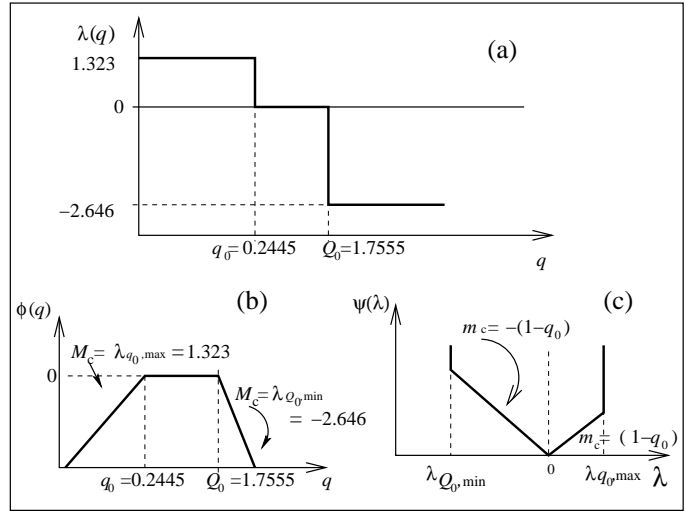}
\caption{$q${\protect\small -phase transitions with index values }$q_{0}$%
{\protect\small \ and }$Q_{0}=2-q_{0}$ {\protect\small obtained from the
main discontinuity in }$\sigma _{n}(m)${\protect\small . See text for
details.}}
\label{fig2}
\end{figure}

We consider the next discontinuities of importance in $\sigma _{n}(m)$.
Independently of the number of discontinuities taken into account one
obtains $q$-exponential forms for $\xi _{t}$. The value of $\sigma
(y=m/2^{n+1})$ at $y=1/4$ measures $1/\alpha _{1}$, with $\alpha _{1}\simeq
5.4588$ for $z=2$, and it is associated to one 'midway' region between the
most crowded and most sparse regions of the attractor (the other 'midway'
region being associated to $\sigma (3/4)$). With three scaling factors, $%
\alpha $, $\alpha _{0}$ and $\alpha _{1}$, we have three values for the $q$
index, $q_{0}$, $q_{1}$ and $q_{2}$ (together with the conjugate values $%
Q_{0}=2-q_{0}$, $Q_{1}=2-q_{1}$ and $Q_{2}=2-q_{2}$ for the inverse
trajectories). To each value of $q$ there is a set of $q$-Lyapunov
coefficients running from a maximum $\lambda _{q_{i},\max }$ to zero (or a
minimum $\lambda _{Q_{i},\min }$ to zero). The results when $z=2$ for the
ranges of values obtained for the $q$-Lyapunov coefficients are shown as the
vertical lines in Figs. 2a, 3a and 4a. Similar results are obtained for the
case of four discontinuities in $\sigma _{n}(m)$, etc.

\section{A family of $q$-phase transitions at the edge of chaos}

As a function of the variable $-\infty <q<\infty $ the $q$-Lyapunov
coefficients obtained in the previous section are functions with two steps
with the jumps located at $q=q_{i}=1-\ln 2/\ln \alpha _{i}(z)/\alpha (z)$
and $q=Q_{i}=2-q_{i}$. Immediate contact can be established with the
formalism developed by Mori and coworkers and the $q$ phase transition
obtained in Ref. \cite{mori2}. Each step function for $\lambda (q)$ can be
integrated to obtain the spectrum $\phi (q)$ ($\lambda (q)\equiv d\phi /dq$)
and from this its Legendre transform $\psi (\lambda )$ ($\equiv \phi
-(1-q)\lambda $). We illustrate this first with $\sigma _{n}(m)$ obtained
with two scale factors, as in Eq. (\ref{sigmafunction2}). We show numerical
values for the case $z=2$. From Eqs. (\ref{q0index}) to (\ref{LyaQ0}) we
obtain 
\begin{equation}
\lambda (q)=\left\{ 
\begin{array}{l}
\lambda _{q_{0},\max },\ q\leq q_{0}=1-\ln 2/(z-1)\ln \alpha \simeq 0.2445,
\\ 
0,\qquad \qquad \qquad \qquad q_{0}<q<Q_{0}, \\ 
\lambda _{Q_{0},\min },\ q\geq Q_{0}=2-q_{0}\simeq 1.7555,
\end{array}
\right.  \label{lambdaq0}
\end{equation}
where $\lambda _{q_{0},\max }=\ln \alpha /\ln 2\simeq 1.323$ and $\lambda
_{Q_{0},\min }=-2\ln \alpha /\ln 2\simeq -2.646$.

The free energy functions $\phi (q)$ and $\psi (\lambda )$ that correspond
to Eq. (\ref{lambdaq0}) are given by 
\[
\phi (q)=\left\{ 
\begin{array}{l}
\lambda _{q_{0},\max }(q-q_{0}),\ q\leq q_{0}, \\ 
0,\qquad q_{0}<q<Q_{0}, \\ 
\lambda _{Q_{0},\min }(q-Q_{0}),\ q\geq Q_{0},
\end{array}
\right. 
\]
and 
\[
\psi (\lambda )=\left\{ 
\begin{array}{l}
(1-Q_{0})\lambda ,\ \lambda _{Q_{0},\min }<\lambda <0, \\ 
(1-q_{0})\lambda ,\ 0<\lambda <\lambda _{q_{0},\max }.
\end{array}
\right. 
\]
We show these functions in Fig. 2. The constant slopes of $\psi (\lambda )$
represent the $q$-phase transitions associated to trajectories linking two
regions of the attractor, that in this case are its most crowded and most
sparse, and their values $1-q_{0}$ and $q_{0}-1$ correspond to those
obtained for the $q$-exponential $\xi _{t}$ Eq. (\ref{sensitivity4}). The
slope $q_{0}-1\simeq -0.7555$ coincides with that originally detected by
Mori and colleagues \cite{mori1}.

When we consider also the next discontinuity of importance in $\sigma _{n}(m)
$, at $\sigma (1/4)=1/\alpha _{1}$, we obtain a set of two $q$-phase
transitions for each of the three values of the $q$ index, $q_{0}$, $q_{1}$
and $q_{2}$. We show in Figs. 2, 3 and 4 the functions $\lambda (q)$, $\phi
(q)$, and $\psi (\lambda )$ obtained for these three cases. The parameter
values for the $q$-phase transitions at $1-q_{0}$ and $q_{0}-1$ appear
again, but now we have also two other sets at $1-q_{1}$ and $q_{1}-1$, and
at $1-q_{2}$ and $q_{2}-1$, that correspond, respectively, to orbits that
link a 'midway' region of the attractor with the most sparse region, and
with the most crowded region of the attractor.

\begin{figure}[htb]
\setlength{\abovecaptionskip}{0pt} \centering
\includegraphics[width=9cm ,angle=0]{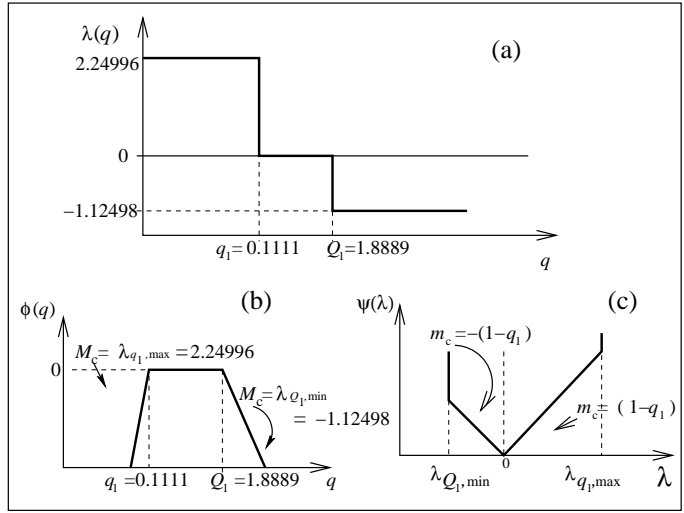}
\caption{$q${\protect\small -phase transitions with index values }$q_{1}$%
{\protect\small \ and }$Q_{1}=2-q_{1}$ {\protect\small obtained from the
main two discontinuities in }$\sigma _{n}(m)${\protect\small . See text for
details.}}
\label{fig3}
\end{figure}

\begin{figure}[htb]
\setlength{\abovecaptionskip}{0pt} 
\centering
\includegraphics[width=9cm 
,angle=0]{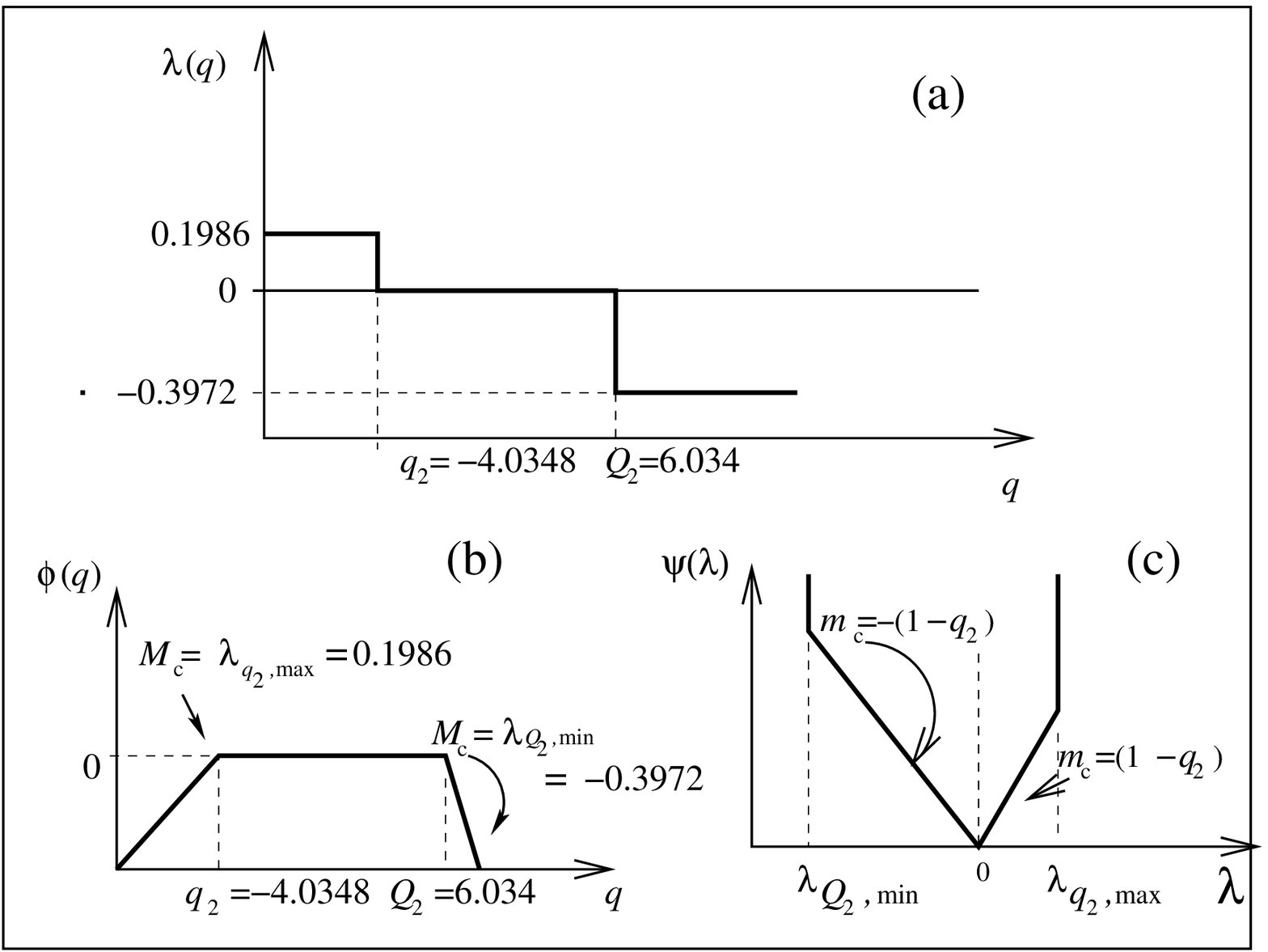}
\caption{$q${\protect\small -phase transitions with index values }$q_{2}$%
{\protect\small \ and }$Q_{2}=2-q_{2}$ {\protect\small obtained from the
main two discontinuities in }$\sigma _{n}(m)${\protect\small . See text for
details.}}
\label{fig4}
\end{figure}

\section{Noisy dynamics at the edge of chaos}

Consider now the logistic map $z$ $=2$ in the presence of additive noise 
\begin{equation}
x_{t+1}=f_{\mu }(x_{t})=1-\mu x_{t}^{2}+\chi _{t}\varepsilon ,\;-1\leq
x_{t}\leq 1,0\leq \mu \leq 2,  \label{logistic1}
\end{equation}
where $\chi _{t}$ is the random variable with average $\left\langle \chi
_{t}\chi _{t^{\prime }}\right\rangle =\delta _{t.t^{\prime }}$, and $%
\varepsilon $ measures the noise intensity \cite{schuster1} \cite
{crutchfield1}. Except for a set of zero measure, all the trajectories with $%
\mu _{\infty }(\varepsilon =0)$ and initial condition $-1\leq x_{0}\leq 1$
fall into the attractor with fractal dimension $d_{f}=0.5338...$. These
trajectories represent nonergodic states, since as $t\rightarrow \infty $
only a Cantor set of positions is accessible out of the total phase space.
For $\varepsilon >0$ the noise fluctuations erase the fine features of the
periodic attractors as these widen into bands similar to those in the
chaotic attractors, yet there remains a definite transition to chaos at $\mu
_{\infty }(\varepsilon )$ where the Lyapunov exponent $\lambda _{1}$ changes
sign. The period doubling of bands ends at a finite value $2^{N(\varepsilon
)}$ as the edge of chaos transition is approached and then decreases at the
other side of the transition. This effect displays scaling features and is
referred to as the bifurcation gap \cite{schuster1} \cite{crutchfield1}.
When $\varepsilon >0$ the trajectories visit sequentially a set of $2^{n}$
disjoint bands or segments leading to a cycle, but the behavior inside each
band is completely chaotic. These trajectories represent ergodic states as
the accessible positions have a fractal dimension equal to the dimension of
phase space. Thus the removal of the noise $\varepsilon \rightarrow 0$ at $%
\mu _{\infty }$ leads to an ergodic to nonergodic transition in the map.

In the absence of noise ($\varepsilon =0$) the diameter positions $%
x_{2^{n}}=d_{n}=$ $\alpha ^{-n}$ visited at times $\tau =2^{n}$ by the
trajectory starting at $x_{0}=1$ is given by the Feigenbaum fixed-point map
solution $g(x)$, 
\begin{equation}
x_{\tau }=\left| g^{^{(\tau )}}(x_{0})\right| =\tau ^{-1/1-q_{0}}\left|
g(\tau ^{1/1-q_{0}}x_{0})\right| ,  \label{trajectory1}
\end{equation}
that in turn is obtained from the $n\rightarrow \infty $ convergence of the $%
2^{n}$th map composition to $(-\alpha )^{-n}g(\alpha ^{n}x)$ with $\alpha
=2^{1/(1-q_{0})}$. When $x_{0}=0$ one obtains in general \cite{robledo2}

\begin{equation}
x_{\tau }=\left| g^{(2l+1)}(0)g^{(2^{n-1})}(0)\right| =\left|
g^{(2l+1)}(0)\right| \alpha ^{-n},\tau =(2l+1)2^{n},\ l,n=0,1,...
\label{trajectory2}
\end{equation}

When the noise is turned on ($\varepsilon $ always small) the $2^{n}$th map
composition converges instead to

\begin{equation}
(-\alpha )^{-n}[g(\alpha ^{n}x)+\chi \varepsilon \kappa ^{n}G_{\Lambda
}(\alpha ^{n}x)],
\end{equation}
where $\kappa $ a constant whose numerically determined \cite{crutchfield2}, 
\cite{shraiman1} value $\kappa \simeq 6.619$ is well approximated by $\nu =2%
\sqrt{2}\alpha (1+1/\alpha ^{2})^{-1/2}$, the ratio of the intensity of
successive subharmonics in the map power spectrum \cite{shraiman1}, \cite
{schuster1}. The connection between $\kappa $ and the $\varepsilon $%
-independent $\nu $ stems from the necessary coincidence of two ratios, that
of noise levels causing band-merging transitions for successive $2^{n}$ and $%
2^{n+1}$ periods and that of spectral peaks at the corresponding parameter
values $\mu _{n}$ and $\mu _{n+1}$ \cite{shraiman1}, \cite{schuster1}.
Following the same procedure as above we see that the orbits $x_{\tau }$ at $%
\mu _{\infty }(\varepsilon )$ satisfy, in place of Eq. (\ref{trajectory1}),
the relation 
\begin{equation}
x_{\tau }=\tau ^{-1/1-q_{0}}\left| g(\tau ^{1/1-q_{0}}x)+\chi \varepsilon
\tau ^{1/1-r}G_{\Lambda }(\tau ^{1/1-q_{0}}x)\right| ,
\end{equation}
where $G_{\Lambda }(x)$ is the first-order perturbation eigenfunction, and
where $r=1-\ln 2/\ln \kappa \simeq 0.6332$. So that use of $x_{0}=0$ yields 
\begin{equation}
x_{\tau }=\tau ^{-1/1-q_{0}}\left| 1+\chi \varepsilon \tau ^{1/1-r}\right|
\end{equation}
or 
\begin{equation}
x_{t}=\exp _{2-q_{0}}(-\lambda _{q_{0}}t)\left[ 1+\chi \varepsilon \exp
_{r}(\lambda _{r}t)\right]
\end{equation}
where $t=\tau -1$, $\lambda _{q_{0}}=$ $\ln \alpha /\ln 2$ ($\lambda
_{q_{0},\max }$ of the previous section) and $\lambda _{r}=\ln \kappa /\ln 2$%
.

At each noise level $\varepsilon $ there is a 'crossover' or 'relaxation'
time $t_{x}=\tau _{x}-1$ when the fluctuations start destroying the detailed
structure imprinted by the attractor on the orbits with $x_{0}=0$. This time
is given by $\tau _{x}=\varepsilon ^{r-1}$, the time when the fluctuation
term in the perturbation expression for $x_{\tau }$ becomes $\varepsilon $%
-independent, i.e. 
\begin{equation}
x_{\tau _{x}}=\tau _{x}^{-1/1-q_{0}}\left| 1+\chi \right| .
\end{equation}
Thus, there are two regimes for time evolution at $\mu _{\infty
}(\varepsilon )$. When $\tau <\tau _{x}$ the fluctuations are smaller than
the distances between adjacent subsequence positions of the noiseless orbit
at $\mu _{\infty }(0)$, and the iterate positions in the presence of noise
fall within small non overlapping bands each around the $\varepsilon =0$
position for that $\tau $. In this regime the dynamics follows in effect the
same subsequence pattern as in the noiseless case. When $\tau \sim \tau _{x}$
the width of the fluctuation-generated band visited at time $\tau _{x}=2^{N}$
matches the distance between two consecutive diameters, $d_{N}-d_{N+1}$
where $N\sim -\ln \varepsilon /\ln \kappa $, and this signals a cutoff in
the advance through the position subsequences. At longer times $\tau >\tau
_{x}$ the orbits are unable to stick to the fine period-doubling structure
of the attractor. In this 2nd regime the iterate follows an increasingly
chaotic trajectory as bands merge progressively. This is the dynamical image
- observed along the \textit{time evolution }for the orbits \textit{of a
single state} $\mu _{\infty }(\varepsilon )$ - of the static bifurcation gap
first described in the map space of position $x$ and control parameter $\mu $
\cite{crutchfield1}, \cite{crutchfield2}.

\section{Analogy with glassy dynamics}

We recall the main dynamical properties displayed by supercooled liquids on
approach to glass formation. One is the growth of a plateau and for that
reason a two-step process of relaxation, as presented by the time evolution
of correlations e.g. the intermediate scattering function $F_{k}$ \cite
{debenedetti1}. This consists of a primary power-law decay in time $t$
(so-called '$\beta $' relaxation) that leads into the plateau, the duration $%
t_{x}$ of which diverges also as a power law of the difference $T-T_{g}$ as
the temperature $T$ decreases to a critical value $T_{g}$. After $t_{x}$
there is a secondary power law decay (so-called '$\alpha $' relaxation) that
leads to a conventional equilibrium state \cite{debenedetti1}. A second
characteristic dynamical property of glasses is the loss of time translation
invariance, a characteristic known as aging \cite{bouchaud1}. The time decay
of relaxation functions and correlations display a scaling dependence on the
ratio $t/t_{w}$ where $t_{w}$ is a waiting time. A third distinctive
property is that the experimentally observed relaxation behavior of
supercooled liquids is described, via regular heat capacity assumptions \cite
{debenedetti1}, by the so-called Adam-Gibbs equation, 
\begin{equation}
t_{x}=A\exp (B/TS_{c}),  \label{adamgibbs0}
\end{equation}
where the relaxation time $t_{x}$ can be identified with the viscosity, and
the configurational entropy $S_{c}$ is related to the number of minima of
the fluid's potential energy surface (and $A$ and $B$ are constants).

Returning to the map, phase space is sampled at noise level $\varepsilon $
by orbits that visit points within the set of $2^{N}$ bands of widths $\sim
\varepsilon $, and this takes place in time in the same way that period
doubling and band merging proceeds in the presence of a bifurcation gap when
the control parameter is run through the interval $0\leq \mu \leq 2$. That
is, the trajectories starting at $x_{0}=0$ duplicate the number of visited
bands at times $\tau =2^{n}$, $n=1,...,N$, the bifurcation gap is reached at 
$\tau _{x}=$ $2^{N}$, after which the orbits fall within bands that merge by
pairs at times $\tau =2^{N+n}$, $n=1,...,N$. The sensitivity to initial
conditions grows as $\xi _{t}=\exp _{q_{0}}(\lambda _{q_{0}}t)$ ($%
q_{0}=1-\ln 2/\ln \alpha <1$) for $t<t_{x}$, but for $t>t_{x}$ the
fluctuations dominate and $\xi _{t}$ grows exponentially as the trajectory
has become chaotic and so one anticipates an exponential $\xi _{t}$ (or $q=1$%
). We interpret this behavior to be the dynamical system analog of the '$%
\alpha $' relaxation in supercooled fluids. The plateau duration $%
t_{x}\rightarrow \infty $ as $\varepsilon \rightarrow 0$. Additionally,
trajectories with initial conditions $x_{0}$ not belonging to the attractor
exhibit an initial relaxation stretch towards the plateau as the orbit falls
into the attractor. This appears as the analog of the '$\beta $' relaxation
in supercooled liquids. See \cite{robglass1}.

Next, we determine the entropy of the orbits starting at $x_{0}=0$ as they
enter the bifurcation gap at $t_{x}(\varepsilon )$ when the maximum number $%
2^{N}$ of bands allowed by the fluctuations is reached. The entropy $%
S_{c}(\mu _{\infty },2^{N})$ associated to the state at $\mu _{\infty
}(\varepsilon )$ has the form 
\begin{equation}
S_{c}(\mu _{\infty },2^{N})=2^{N}\varepsilon s,
\end{equation}
since each of the $2^{N}$ bands contributes with an entropy $\varepsilon s$
where 
\begin{equation}
s=-\int_{-1}^{1}p(\chi )\ln p(\chi )d\chi ,
\end{equation}
and where $p(\chi )$ is the distribution for the noise random variable. In
terms of $t_{x}$, given that $2^{N}=$ $1+t_{x}$ and $\varepsilon
=(1+t_{x})^{-1/1-r}$, one has 
\begin{equation}
S_{c}(\mu _{\infty },t_{x})/s=(1+t_{x})^{-r/1-r}
\end{equation}
or, conversely, 
\begin{equation}
t_{x}=(s/S_{c})^{(1-r)/r}.  \label{adamgibbs1}
\end{equation}
Since $t_{x}\simeq \varepsilon ^{r-1}$, $r-1\simeq -0.3668$ and $%
(1-r)/r\simeq 0.5792$ then $t_{x}\rightarrow \infty $ and $S_{c}\rightarrow
0 $ as $\varepsilon \rightarrow 0$, i.e. the relaxation time diverges as the
'landscape' entropy vanishes. We interpret this relationship between $t_{x}$
and the entropy $S_{c}$ to be the dynamical system analog of the Adam-Gibbs
formula for a supercooled liquid. See \cite{robglass1}. Notice that Eq.(\ref
{adamgibbs1}) is a power law in $S_{c}^{-1}$ while for structural glasses it
is an exponential in $S_{c}^{-1}$ \cite{debenedetti1}.

Last, we examine the aging scaling property of the trajectories $x_{t}$ at $%
\mu _{\infty }(\varepsilon )$. The case $\varepsilon =0$ is more readily
appraised because this property is actually built into the position
subsequences $x_{\tau }=\left| g^{(\tau )}(0)\right| $, $\tau =(2l+1)2^{n}$, 
$l,n=0,1,..$. These subsequences are relevant for the description of
trajectories that are 'detained' at a given attractor position for a waiting
period of time $t_{w}$ and then 'released' to the normal iterative
procedure. We chose the holding positions to be any of those along the top
band shown in Fig. 1 for a waiting time $t_{w}=2l+1$, $l=0,1,...$. Notice
that, as shown in Fig. 1, for the $x_{0}=0$ orbit these positions are
visited at odd iteration times. The lower-bound positions for these
trajectories are given by those of the subsequences at times $(2l+1)2^{n}$
(see Fig. 1). Writing $\tau $ as $\tau =$ $t_{w}+t$ we have that $%
t/t_{w}=2^{n}-1$ and 
\begin{equation}
x_{t+t_{w}}=g^{(t_{w})}(0)g^{(t/t_{w})}(0)
\end{equation}
or 
\begin{equation}
x_{t+t_{w}}=g^{(t_{w})}(0)\exp _{q_{0}}(-\lambda _{q_{0}}t/t_{w}).
\label{trajectory5}
\end{equation}
This property is gradually modified when noise is turned on. The presence of
a bifurcation gap limits its range of validity to total times $t_{w}+t$ $%
<t_{x}(\varepsilon )$ and so progressively disappears as $\varepsilon $ is
increased. See Ref. \cite{robglass1}.

\section{Concluding remarks}

We have re-examined the dynamical properties at the onset of chaos in the
logistic map and obtained further understanding about their nature. We
exhibited links between original developments, such as Feigenbaum's $\sigma $
function, Mori's $q$-phase transitions and the noise-induced bifurcation
gap, with more recent advances, such as $q$-exponential sensitivity to
initial conditions \cite{robledo2}, $q$-generalized Pesin identity \cite
{robledo4}, and dynamics of glass formation \cite{robglass1}. The dynamics
at the edge of chaos is anomalous because it is an incipient chaotic
attractor with vanishing ordinary Lyapunov coefficient $\lambda _{1}$.
Chaotic orbits possess a time irreversible property that stems from mixing
in phase space and loss of memory, but orbits within critical attractors are
non-mixing and have no loss of memory. As a classic illustration of the
latter case the attractor at the onset of chaos presents dynamical
properties with self-similar structure that result in a set of power laws
for the sensitivity to initial conditions. We determined exact analytical
expressions for $\xi _{t}$.

Our most striking finding is that the dynamics at the onset of chaos is
constituted by an infinite family of Mori's $q$-phase transitions, each
associated to orbits that have common starting and finishing positions
located at specific regions of the attractor. Each of these transitions is
related to a discontinuity in the $\sigma $ function of 'diameter ratios',
and this in turn implies a $q$-exponential $\xi _{t}$ and a spectrum of $q$%
-Lyapunov coefficients for each set of orbits. The transitions come in pairs
with specific conjugate indexes $q$ and $Q=2-q$, as these correspond to
switching starting and finishing orbital positions. Since the amplitude of
the discontinuities in $\sigma $ diminishes rapidly, in practical terms
there is only need of evaluation for the first few of them. The dominant
discontinuity is associated to the most crowded and sparse regions of the
attractor and this alone provides a very reasonable description, as found in
earlier studies \cite{tsallis2} \cite{tsallis3} \cite{robledo2} \cite
{robledo4}. Thus, the special values for the Tsallis entropic index $q$ in $%
\xi _{t}$ are equal to the special values of the variable $q$ in the
formalism of Mori and colleagues at which the $q$-phase transitions take
place.

As described, the dynamics of noise-perturbed logistic maps at the chaos
threshold presents the characteristic features of glassy dynamics observed
in supercooled liquids. In particular our results are \cite{robglass1}: i)
The two-step relaxation that takes place when $\varepsilon \rightarrow 0$ is
obtained in terms of the bifurcation gap properties, specifically, the
plateau duration $t_{x}$ is given by a power law in the noise amplitude $%
\varepsilon $. ii) The map analog of the Adam-Gibbs law is given also as a
power-law relation between $t_{x}(\varepsilon )$ and the entropy $%
S_{c}(\varepsilon )$ associated to the noise widening of chaotic bands. iii)
The trajectories at $\mu _{\infty }(\varepsilon \rightarrow 0)$ are shown to
obey a scaling property, characteristic of aging in glassy dynamics, of the
form $x_{t+t_{w}}=h(t_{w})h($ $t/t_{w})$ where $t_{w}$ is a waiting time.

The limit of vanishing noise amplitude $\varepsilon \rightarrow 0$ (the
counterpart of the limit $T-T_{g}\rightarrow 0$ in the supercooled liquid)
brings about loss of ergodicity. This nonergodic state with $\lambda _{1}=0$
corresponds to the limiting state, $\varepsilon \rightarrow 0$, $%
t_{x}\rightarrow \infty $, for a family of small $\varepsilon $ states with
glassy properties, which are expressed for $t<t_{x}$ via the $q$%
-exponentials of the Tsallis formalism. It has been suggested on several
occasions \cite{tsallis4} \cite{tsallis5} that the setting in which
nonextensive statistics appears to come out is linked to the prevalence of
nonuniform convergence, such as that involving the thermodynamic $%
N\rightarrow \infty $ and the infinitely large time $t\rightarrow \infty $
limits. Here a similar situation happens, that is, if $\varepsilon
\rightarrow 0$ is taken before $t\rightarrow \infty $ a nonergodic orbit
restrained to the Feigenbaum attractor and with fully-developed glassy
properties is obtained, whereas if $t\rightarrow \infty $ is taken before $%
\varepsilon \rightarrow 0$ a chaotic orbit with $q=1$ would be observed.

\textbf{Acknowledgments}. AR is grateful for the support provided by the
Verhulst 200 organizers. We also acknowledge support from CONACyT and
DGAPA-UNAM, Mexican agencies.

\end{document}